\documentclass[twocolumn,showpacs,preprintnumbers,amsmath]{revtex4}

\usepackage{graphicx}
\usepackage{dcolumn}
\usepackage{bm}
\begin{document}

\title{Quantum phase transition in the one-dimensional period-two and
uniform  compass model}
\author{Ke-Wei Sun$^{1}$, Yu-Yu Zhang$^{1}$, and Qing-Hu Chen$^{2,1,*}$}
\address{$^{1}$ Department of Physics, Zhejiang University, Hangzhou 310027,
P. R. China \\
$^{2}$ Center for Statistical and Theoretical Condensed Matter
Physics, Zhejiang Normal University, Jinhua 321004, P. R. China}
\date{\today}
\begin{abstract}
Quantum phase transitions in the one-dimensional period-two and
uniform quantum compass model are studied by using the pseudo-spin
transformation method and the trace map method. The exact solutions
are presented, the fidelity, the nearest-neighbor pseudo-spin
entanglement, spin and pseudo-spin correlation functions are then
calculated. At the critical point, the fidelity and its
susceptibility change substantially, the gap of pseudo-spin
concurrence is observed, which scales as $1/N$ (N is the system
size). The spin correlation functions show smooth behavior around
the critical point. In the period-two chain, the pseudo-spin
correlation functions exhibit an oscillating behavior, which is
absent in the uniform chain. The divergent correlation length at the
critical point is demonstrated in the general trend for both cases.
\end{abstract}


\maketitle

\section{introduction}

Recently, the quantum compass model was introduced to describe some
Mott insulators with orbit degeneracy by a
pseudospin\cite{Doucot,Mishra}, where the coupling along one of
bonds is an Ising type, but different spin components are active
along other bond directions. The disorder effect in this model was
also examined\cite{Tanaka}. The protected qubit is formed if it is
separated from the low-energy excitations by a pseudo-spin excited
gap. So a high quality factor, scalable and error-free scheme of
quantum computation can be designed\cite{HDChen}. The symmetry of
pseudo-spin Hamiltonians is usually much lower than
SU(2)\cite{Brink}, and the result of numerical calculation has been
shown that its eigenstates are at least twofold degenerate or highly
degenerate and disordered\cite{Dorier}. The quantum XX-ZZ model,
also called one-dimensional (1D) compass model, is constructed by
antiferromagnetic order of X and Z pseudo-spin components on odd and
even bonds, respectively\cite{Brzezicki}.  In addition, the 1D
quantum compass model is exactly the same as the 1D reduced Kitaev
model\cite{Feng}.  The analytic eigenspectra  in the latter model
have been obtained, and it was shown  that this  model has one
gapless phase. But the characters of the quantum phase transition
have never been well studied  previously. The realistic models of
the orbital degeneracy are more complicated.

For the compass model, the pseudo-spins may lead to enhanced quantum
fluctuations near the quantum phase transitions (QPTs) and to
entangled spin-orbital ground states. The numerical results have
indicated that a first-order QPT occurs at $J_x=J_z$ between two
different states with spin ordering along either x or z
directions\cite{Dorier}. Recently, the ground-state (GS) fidelity
\cite{Quan,You,Buonsante,Cozzini,Chen,zhou,Liu} and entanglement
\cite{Osterloh,Tong,Zhang,Emary,Liberti,Reslen,chenqh,Osborne,Tong2}
emerged from quantum information science have been used in signaling
the QPTs.  To calculate these quantities accurately, it is necessary
to know the exact GS wave function. The derivatives of the GS energy
are intrinsically related to the GS fidelity\cite{Chen}, both can be
used to identify the QPTs. For the special case of two spin $-1/2$
system, the entanglement is given by the concurrence. Quantum
entanglement is one of the most striking consequences of quantum
correlation in many-body systems, shows a deep relation with the
QPT\cite {Osterloh}. Therefore understanding the entanglement is
very important in QPTs\cite{Tong,Zhang}. In the context of QPTs, the
quantum entanglement have been the subject of considerable interests
in the Dicke model \cite{Emary,Liberti,chenqh} and the XY model
\cite{Osborne,Reslen}.

On the other hand, experimental works on quasicrystals
\cite{Shechtman} and quasiperiodic superlattices \cite{Merlin} have
inspired theoretical interests in 1D quasiperiodic systems.
Period-two chain can be regarded as the intermediate one between
uniform periodic chain and quasiperiodic chain, which have exhibited
some unusual physical properties. In this work, we study the
one-dimensional compass model for both uniform and period-two cases
by using transfer matrix method\cite{Tong2} and the method of Lieb,
Schultz, and Mattis\cite{Lieb}. The exact solutions for two cases
are obtained. The GS fidelity and the energy gap between uniform and
period-two quantum spin chain are calculated. The behaviors of the
pseudo-spin correlations with periodic boundary condition are given.

The paper is organized as follows: In Section \textbf{II}, we give
the model and the exact solution with periodic boundary condition.
The calculation methods of fidelity and concurrence are introduced
in Section \textbf{III}. The correlation functions are analyzed in
Section \textbf{IV}. The paper is summarized in Section \textbf{V},
where we give some discussions and conclusions.

\section{MODEL HAMILTONIAN AND EXACT SOLUTION}

The  Hamiltonian of one-dimensional  compass model is given by
\begin{equation}
H=\sum_{i=1}^{N^{\prime }}[J_i(\sigma _{2i-1}^z\sigma
_{2i}^z+\beta \sigma _{2i}^x\sigma _{2i+1}^x)],
\end{equation}
where $J_i$ is the nearest-neighbor interaction, $\sigma _i^{x(z)}
$ are the Pauli matrix on site $i$, $N=2N^{\prime }$ is the number
of the sites, and $\beta $ is the coupling parameter which
determines the phase transition point. For $J_{2i}=J$ and
$J_{2i+1}=\alpha J$ , the model is a period-two case. By using the
pseudo-spin (orbital) transformation method which is given by
Brzezicki \emph{et al}\cite{Brzezicki}, we can define the
modulated interactions for odd pairs of pseudo-spins $\{2i-1,2i\}$
as $-\tau _i^z\equiv \sigma _{2i-1}^z\sigma _{2i}^z$, and the
spin-flip operators of $x$ direction are given by
$\tau_{i}^{x}\equiv (-1)^{\sum_{k=1}^{i-1}s_k} \prod_{j=
2i}^{2N^{\prime}}\sigma_{j}^{x}$. The two neighboring odd bonds
can be expressed as the even $\{2i,2i+1\}$ bonds by a product
$-\tau _i^x\tau _{i+1}^x$. Then the Hamiltonian of one-dimensional
compass model can be written as follows
\begin{eqnarray}
H_{\vec{s}} &=&-\sum_{i=1}^{N^{\prime }-1}[J_i(\tau _i^z+\beta
\tau _i^x\tau
_{i+1}^x)]  \nonumber \\
&&-J_{N^{\prime }}[\tau _{N^{\prime }}^z+(-1)^s\beta \tau
_{N^{\prime }}^x\tau _1^x].
\end{eqnarray}
Note that it looks like but is different from the transverse field
Ising model.

The vector $\vec{s}$ represents the state $(s_1,\cdots ,s_{N^{\prime
}})$.
Here $s_i=1$ ($s_i=0$) labels that the two pseudo-spins of the odd bond $\{%
2i-1,2i\}$ are parallel (antiparallel). $s=\sum_{i=1}^{N^{\prime
}}s_i$ is the number of parallel odd pairs of spins. In this paper,
we only discuss the ferromagnetic boundary condition of the quantum
compass model, i.e. the case of the even $s$. The effective
Hamiltonian (2) can be solved by using the Jordan-Wigner
transformation for spin operators,
\begin{equation}
\tau _i^z=1-2c_i^{\dagger }c_i,
\end{equation}
\begin{equation}
\tau _i^x=(c_i+c_i^{\dagger })\prod_{j<i}(1-2c_j^{\dagger }c_j),
\end{equation}
where $c_i$ and $c_i^{\dagger }$ are the anticommuting fermion
operators. After this transformation, The effective Hamiltonian
becomes
\begin{eqnarray}
H_{\vec{s}} &=&\sum_{i=1}^{N^{\prime }-1}[2J_ic_i^{\dagger }c_i+  \nonumber
\\
&&J_i\beta (c_ic_{i+1}^{\dagger }-c_i^{\dagger }c_{i+1}^{\dagger
}+c_ic_{i+1}-c_i^{\dagger }c_{i+1})]  \nonumber \\
&&+J_{N^{\prime }}\beta (c_{N^{\prime }}\tilde{c}_1^{\dagger}
-c_{N^{\prime }}^{\dagger }\tilde{c}_1^{\dagger }+c_{N^{\prime
}}\tilde{c}_1-c_{N^{\prime
}}^{\dagger }\tilde{c}_1)  \nonumber \\
&&+J_{N^{\prime }}c_{N^{\prime }}^{\dagger }c_{N^{\prime
}}-\sum_{i=1}^{N^{\prime }}J_i,
\end{eqnarray}
with
\begin{equation}
\tilde{c}_1=c_1(-1)^{1+s+\sum_{j=1}^{N^{\prime }}c_j^{\dagger }c_j}.
\end{equation}

Because we assume that the parity of $s$ is even, it implies that
only states with even numbers of Bogoliubov quasiparticles in the
spectrum of the Hamiltonian (5). Under the  periodic boundary
condition $(c_{N^{\prime }+1}=c_1)$, the number of $c$ fermions must
be odd parity, as can easily be obtained from equation (6). Then the
general form of the Hamiltonian is simplified to
\begin{eqnarray}
H_{\vec{s}} &=&\sum_{i=1}^{N^{\prime }}[2J_ic_i^{\dagger }c_i+J_i\beta
(c_ic_{i+1}^{\dagger }-c_i^{\dagger }c_{i+1}^{\dagger
}+c_ic_{i+1}-c_i^{\dagger }c_{i+1})]  \nonumber \\
&&-\sum_{i=1}^{N^{\prime }}J_i.
\end{eqnarray}
For the period-two case, we can rewrite (7) as the following form by
neglecting the last constant term,
\begin{equation}
H=\sum_{i,j=1}^{N^{\prime }}[c_i^{\dagger }A_{ij}c_j+\frac 12(c_i^{\dagger
}B_{ij}c_j^{\dagger }+h.c.)],
\end{equation}
where the nonzero elements of the matrices \textbf{A} and \textbf{B} are
given by
\begin{eqnarray*}
A_{ij} &=&2J_i\delta _{i,j}-J_i\beta \delta _{j,i+1}-J_j\beta
\delta_{j,i-1}, \\
B_{ij} &=&-J_i\beta \delta _{j,i+1}+J_j\beta \delta _{j,i-1}; \\
A_{1N^{\prime }} &=&A_{N^{\prime }1}=-J_{N^{\prime }}\beta, \\
B_{1N^{\prime }} &=&-B_{N^{\prime }1}=J_{N^{\prime }}\beta.
\end{eqnarray*}
Equation (8) can be diagonalized by using the Bogoliubov
transformation
\begin{eqnarray}
\eta _k &=&\frac 12\sum_{i=1}^{N^{\prime }}[(\phi
_{ki}+\psi_{ki})c_i+(\phi
_{ki}-\psi_{ki})c_i^{\dagger }],  \nonumber \\
\eta _k^{\dagger } &=&\frac 12\sum_{i=1}^{N^{\prime }}[(\phi _{ki}+\psi_{ki}%
)c_i^{\dagger }+(\phi _{ki}-\psi_{ki})c_i],
\end{eqnarray}
where $\psi _{ki}$ is the eigenvector of the matrix $(\textbf{A}+\textbf{B})(%
\textbf{A}-\textbf{B})$ and $\phi _{ki}$ is that of the matrix  $(\textbf{A}-%
\textbf{B})(\textbf{A}+\textbf{B})$. The eigenvalues of both
matrices are corresponding to $\Lambda _k^2$. We take
$k=0,\pm \frac{2\pi }{N^{\prime }},\pm 2\frac{2\pi }{N^{\prime }}%
,\ldots ,\pi $. This relation is satisfied with the periodic
boundary condition. In general, the two eigenvectors ($\phi _{ki}$
and $\psi _{ki}$) satisfy the following equations
\begin{equation}
(A-B)\vec{\psi _k}=\Lambda _k\vec{\phi} _k, (A+B)\vec{\phi} _k=\Lambda _k\vec{%
\psi} _k,
\end{equation}
where $\vec{\phi} _k$ and $\vec{\psi} _k$ are two column vectors.
The diagonalized result takes the form
\begin{equation}
H=\sum_k\Lambda _k(\eta _k^{\dagger }\eta _k-\frac 12).
\end{equation}
The excitation energies $\Lambda _k \ge 0$. At zero temperature,
the QPT points are those parameters that satisfy the condition
$\Lambda _k=0$, and the two coupled coefficients of the Bogoliubov
transformation satisfy the following equations:
\begin{eqnarray}
\Lambda _k\phi _{k,i} &=&2J_i\psi _{k,i}-2J_{i-1}\beta \psi _{k,i-1},\nonumber \\
\Lambda _k\psi _{k,i} &=&2J_i\beta
\phi_{k,i}-2J_i\beta\phi_{k,i+1},
\end{eqnarray}
which can be derived from equation (10). For the period-two case,
i.e. $J_{2i}=J$ and $J_{2i+1}=J\alpha $, if we take $J=1$ and assume
that $\psi _{k,2n}=Ae^{i2nk}$ and $ \psi _{k,2n+1}=Be^{i(2n+1)k}$,
the exact results of $\Lambda _k$ can be obtained analytically from
the coupled equations (12) by using the trace map method. The result
is expressed as
\begin{eqnarray}
\Lambda _{k\pm }^2 &=&\pm \sqrt{4J^4(\beta ^2+1)^2(\alpha ^2-1)^2+64\alpha
^2J^4\beta ^2\cos ^2k}  \nonumber \\
&&+(\alpha ^2+1)(2J^2\beta ^2+2J^2).
\end{eqnarray}
The excitation energies have two branches ($\Lambda _{k-}$ and
$\Lambda _{k+} $). For a special case $\alpha=1$, i.e., the uniform
periodic chain, the excitation energies can be simplified as
$2J\sqrt{1+\beta^2-2\beta\cos k}$, which is the same as that in Ref.
\cite{Brzezicki}. The QPT point is determined by $\Lambda _{k-}$. At
the critical point, the equation can be decoupled for $\Lambda_k=0$.
One of equation (12) is rewritten as $\phi _{k,i+1}=\frac 1\beta
\phi _{k,i}$. Due to the periodic boundary condition,  $(\frac
1\beta )^{N^{\prime }}=1$ should be satisfied. The only possibility
is $\beta =1$, i.e., there is only one QPT point at $\beta =1$ in
this case. On the other hand, the GS energy is expressed as
$E_0=-\frac 12\sum_k\Lambda _k$ which includes the spectra of the
$\pm $ branches. In thermodynamic limit, the summation can be
replaced by an integral
\begin{equation}
E_0=-JN^{\prime }\frac 1{2\pi }\int_0^\pi (\Lambda _{k-}+\Lambda _{k+})dk.
\end{equation}
The pseudo-spin excitation gap $\Delta$, which is energy
difference between the first excited state and the ground state,
is equal to $\Lambda _{0-}$, which disappears at $\beta =1$.

\begin{figure}[tbp]
\includegraphics[scale=0.8]{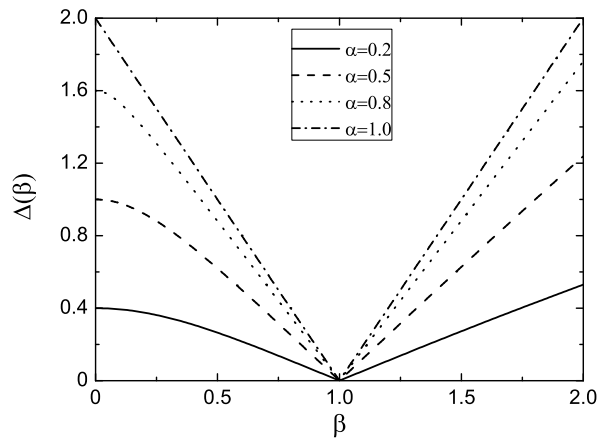}
\caption{Pseudo-spin excitation gap $\Delta $ on uniform and
period-two cases of the compass model. The gaps collapse at the
quantum phase transition point at $\beta =1$ for different values of
$\alpha$. } \label{gap}
\end{figure}

From the Fig.1, we can find that the symmetries of the pseudo-spin
gaps are broken more obviously as $\beta $ is away from the QPT
point in the period-two model. The symmetries remain for the
uniform model. The quantum critical point is fixed at $\beta_c=1$
which separates the disorder phase. In the vicinity of the quantum
critical point, the linear relation $\Delta =\sqrt{10(\alpha
^2+1)-2\sqrt{25\alpha ^4+14\alpha ^2+25}}|1-\beta|$ is generally
satisfied.

\section{FIDELITY AND PSEUDO SPIN CONCURRENCE}

The exact GS wave function of the system must be obtained in order
to calculate the fidelity and concurrence. Similar to the Bardeen,
Cooper, and  Schrieffer GS wave function, we can write the present
GS wavefunction as\cite{Chung}:
\begin{equation}
|\Psi _0(\beta )\rangle =\prod_k\eta _k|Vac\rangle\quad\mbox{for
all k}.
\end{equation}
According to equation (9) and the definition of the
fidelity\cite{Buonsante}
\begin{equation}
F(\beta ,\delta )=|\langle \Psi _0(\beta )|\Psi _0(\beta +\delta )\rangle |,
\end{equation}
where $\delta$ is a small quantity ($\delta=10^{-4}$ is taken in our
calculation), the fidelity and its susceptibility can be given by
\begin{eqnarray}
F(\beta ,\delta ) &=&\prod_k|\sum_i\frac 14[\phi _{ki}(\beta)-\psi
_{ki}(\beta)][\phi_{ki}(\beta+\delta)\nonumber \\
&&-\psi _{ki}(\beta+\delta)]|, \\
S(\beta ) &=&2\lim_{\delta \to 0}\frac{1-F(\beta ,\delta
)}{\delta^2}.
\end{eqnarray}

The numerical results for the GS fidelity and its susceptibility are
plotted in Fig. 2. An abrupt jump occurs in the vicinity of the QPT
point ($\beta_c=1$) as a consequence of the dramatic change of the
structure of the GS. It agrees exactly with our analytical
derivations. One can see level-crossing at $\beta=\beta_c$,
indicating the first-order QPT in this model.

\begin{figure}[tbp]
\includegraphics[scale=0.7]{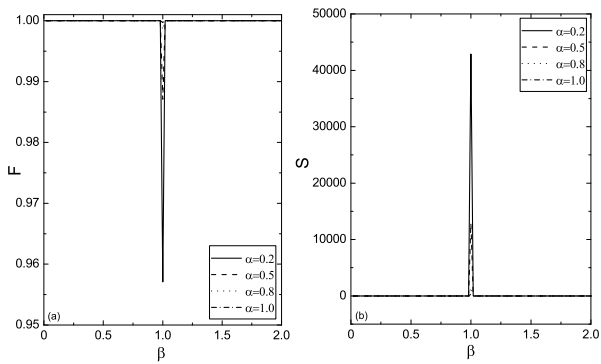}
\caption{The fidelity and the susceptibility of the period-two
compass model versus $\beta$ for $\alpha=0.2,0.5,0.8,1.0$ and
$N^{\prime}=100$. The first QPT point is obviously found at
$\beta_c=1$.} \label{periodic boundary fidelity}
\end{figure}

In recent years, the concept of concurrence is usually adopted as
the measure of the entanglement in spin $-\frac 12$ systems. We
will give the nearest-neighbor pseudo-spin two-point correlation
functions to calculate the nearest-neighbor concurrence (NNC) of
the system. Because of the reflection symmetry, the global phase
flip symmetry, and the Hamiltonian being real, the nonzero
elements are given by \cite{Zhang,Lieb}
\begin{eqnarray}
\langle \tau _i^x\tau _{i+1}^x\rangle  &=&G_{i,i+1}, \langle \tau
_i^y\tau_{i+1}^y\rangle =G_{i+1,i},  \nonumber \\
\langle \tau _i^z\tau _{i+1}^z\rangle
&=&G_{i,i}G_{i+1,i+1}-G_{i,i+1}G_{i+1,i},  \nonumber \\
\langle \tau _i^z\rangle  &=&-G_{i,i},
\end{eqnarray}
where $G_{i,j}=-\sum_k\psi _{ki}\phi _{kj}$. The definition of
concurrence is given by
$C(i,j)=max[r_1(i,j)-r_2(i,j)-r_3(i,j)-r_4(i,j),0]$, where $
r_\alpha (i,j)$ are the square roots of the eigenvalues of the
product matrix $R=\rho _{ij}\tilde{\rho _{ij}}$ in descending order.
The spin flipped matrix $\tilde{\rho _{ij}}$ is defined as
$\tilde{\rho _{ij}}=(\sigma ^y\otimes \sigma ^y)\rho_{ij}^{*}(\sigma
^y\otimes \sigma ^y)$. The $\rho _{ij}$ is the density matrix for a
pair of qubits from a multi-qubit state. In this way,  we can
calculate the NNC of pseudo-spins. For the period-two chain, the
concurrence $C_{2i,2i+1}$ and $C_{2i+1,2i+2}$ are different. So we
use the average concurrence $C=\frac 12(C_{2i,2i+1}+C_{2i+1,2i+2})$.

\begin{figure}[tbp]
\includegraphics[scale=0.7]{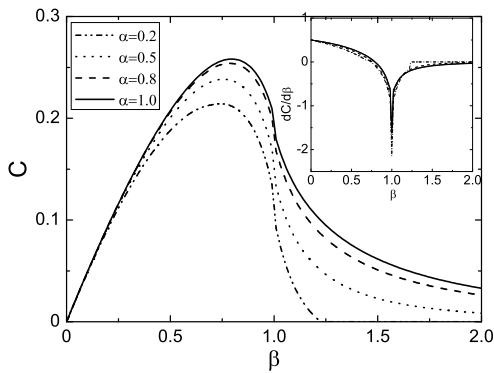}
\caption{The concurrence $C$   versus $\beta$ for
$\alpha=0.2,0.5,0.8,1.0$ ($N^{\prime}=100$).  The inset shows the
derivative $\partial_{\beta}C$ as a function of $\beta$. }
\label{con1}
\end{figure}

The numerical results for the concurrence as a function of $\beta$
are given in Fig. 3. It is shown that the maximum value of the
concurrence gradually increases with the increase of parameter
$\alpha$. If $\alpha$ is small enough, the entanglement of
nearest-neighbor pseudo-spins disappears in the larger $\beta$
regime. A cusp of the first derivative of the concurrence occurs
at the critical point $\beta=1$, similar to those in Ref.
\cite{Osterloh}.

\begin{figure}[tbp]
\includegraphics[scale=0.7]{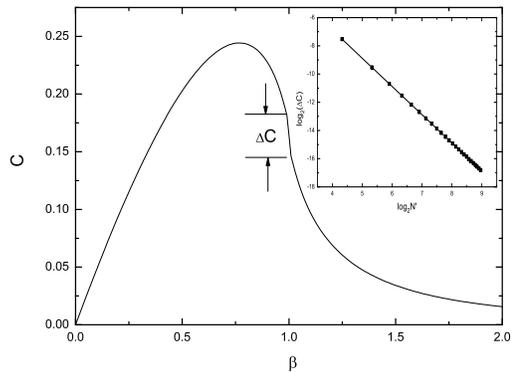}
\caption{The NNC as a function of $\beta$ with $\alpha=0.6$ for $
N^{\prime}=100$. A gap $\Delta C$ for the concurrence is found at
the critical point. The inset shows the size scaling of the gap.}
\label{conN100R and congap scaling}
\end{figure}

A gap is found in the curve of NNC versus $\beta$ at the QPT point
$\beta_c$ in our calculation of the pseudo-spin concurrence. If the
pseudo-spin chain goes to infinite, the gap has the critical
behavior with $\Delta C\propto N^{-1}$, as shown in the inset of
Fig.4. Obviously, it is the finite-size effect. The question then
arises: what is the origin of the concurrence gap? The answer is the
symmetry of system which has been assumed by the ferromagnetic
even-pseudo-spin chains with periodic boundary condition in this
paper. The QPT in the 1D compass model is of
first-order\cite{HDChen,Brzezicki}, the scaling behaviors  at the
critical point should be absent. But the discontinuousness of the
concurrence at the QTP may exhibit the finite-size scaling behavior
$N\to-1$\cite{Liaw}, consistent with the present observation. Due to
the concurrence gap, the value of $\partial_{\beta}C$ becomes
minimum at the critical point. However, the maximum value of the
concurrence occurs below $\beta_c$ is not related to the critical
point. The present results for the concurrence  are similar to those
in the periodic quantum Ising chain model\cite{Zhang}.

\section{SPIN AND PSEUDO-SPIN CORRELATION FUNCTIONS}

Firstly, we show the numerical results of the ground-state spin
correlations on odd $\{2i-1,2i\}$ and even $\{2i,2i+1\}$ bonds as a
function of $\beta$ with a periodic boundary condition. The value of
$<\sigma_{2i-1}^z\sigma_{2i}^z>$ gradually increases with $\beta$
while $<\sigma_{2i}^x\sigma_{2i+1}^x>$ decreases with $\beta$, as
shown in Fig. 5. The crossing points of
$<\sigma_{2i-1}^z\sigma_{2i}^z>$ and
$<\sigma_{2i}^x\sigma_{2i+1}^x>$ curves for  the same $\alpha$ occur
at the quantum critical point.  Actually, the compass model is a
kind of pseudo-spin Ising chains at $\beta=0$ and $\beta\to\infty$.
As a result, the curves of spin correlations versus $\beta$ are
asymmetric. So $<\sigma_{2i-1}^z\sigma_{2i}^z>\to 0$ and
$<\sigma_{2i}^x\sigma_{2i+1}^x>\to -1$ as $\beta\to\infty$.

\
\begin{figure}[tbp]
\includegraphics[scale=0.7]{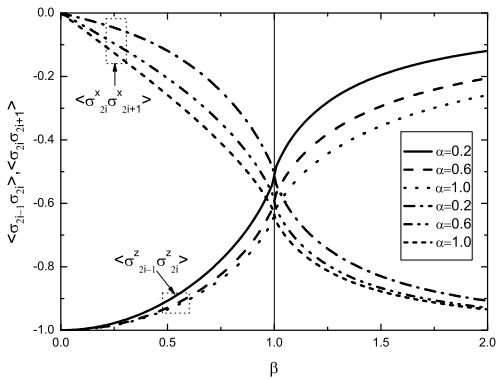}
\caption{Spin correlation functions in the period-two chain and
uniform chain for $\alpha=0.2,0.6,1.0$ and $N^{\prime}=100$. }
\label{spin correlation}
\end{figure}

It is found that the correlation gradually increases with the
decreasing $\alpha$. When $\alpha=1$, the numerical result at the
critical point is the same as the analytical result by Brzezicki
et al. \cite{Brzezicki}.

\begin{figure}[tbp]
\includegraphics[scale=0.7]{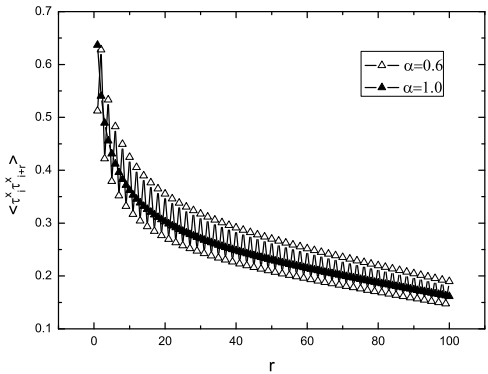}
\caption{Distance dependence of $<\tau_{i}^x\tau_{i+r}^x>$
correlator at the critical point. The parameters are
$\alpha=0.6,1.0$ and $N^{\prime}=200$.} \label{periodic boundary
condition deltacon scaling}
\end{figure}

Finally, we calculate the distance dependence of the pseudo-spin
correlator $<\tau_{i}^x\tau_{i+r}^x>$ under the periodic boundary
condition for the period-two and uniform cases. The the two-point
correlation function is given by\cite{Lieb}

\begin{equation}
<\tau _i^x\tau _{i+r}^x>=\left|
\begin{array}{llll}
G_{i,i+1} &G_{i,i+2} & ... & G_{i,i+r} \\
G_{i+1,i+1} & G_{i+1,i+2} & ... & G_{i+1,i+r}\\
... & ... & ... & ... \\
G_{i+r-1,i+1} & G_{i+r-1,i+2} & ... & G_{i+r-1,i+r}
\end{array}
\right| ,
\end{equation}

which has the form of Toeplitz determinant. When $r\to\infty$, the
correlators gradually decrease and approach the asymptotic value
for large $r$ in an algebraic way\cite{Brzezicki}. This correlator
is positive for all $r$, indicating that there is the long-range
ferromagnetic order. It is interesting to find that the
oscillation occurs for $\alpha\neq 1$, i.e. for period-two chain,
which can be attributed to the different coupling coefficients of
odd and even bonds. However, the similar trend appears in both
cases, as shown in Fig. 6.

\section{SUMMARY and DISCUSSION}

By using the pseudo-spin transformation method and the trace map
method, we obtain the exact solution of one-dimensional compass
model with periodic boundary condition. The parameter $\alpha$
determines the symmetries of finite pseudo-spin excitation gap
$\Delta$, but the phase transition point is still fixed at
$\beta=1$. The quantum critical point separates the disorder
phase. The pseudo-spin liquid disordered ground state is the
universal features in the 1D compass model. The numerical methods
to calculate the fidelity and concurrence are also given. We
observe a first-order quantum phase transition between two
different disordered phase. The concurrence gap $\Delta C$
displays the scaling property $N=-1$. The spin and pseudo-spin
correlation functions are calculated. Curves for the two spin
correlation function cross exactly at the critical point for any
value of $\alpha$. It is observed that the distance dependence of
$<\tau_{i}^x\tau_{i+r}^x>$ correlator displays oscillation in the
period-two case, and a divergent correlation length at the
critical point is observed in both uniform and period-two chains.

\section*{ACKNOWLEDGEMENTS}

We acknowledge useful discussions with  Peiqing Tong and Guang-Ming
Zhang. This work was supported by National Natural Science
Foundation of China, PCSIRT (Grant No. IRT0754) in University in
China, National Basic Research Program of China (Grant No.
2009CB929104), Zhejiang Provincial Natural Science Foundation under
Grant No. Z7080203, and Program for Innovative Research  Team  in
Zhejiang Normal University.

$^{*}$ Corresponding author. Email:qhchen@zju.edu.cn

\end{document}